\documentstyle[prb,aps,epsf]{revtex} \def\narrowtext{} \tighten \twocolumn
\input epsf.sty
 
\begin{document}
\draft
 
\title{Extraction of the Electron Self-Energy from Angle Resolved
Photoemission Data:  Application to Bi2212}
\author{M. R. Norman$^1$, H. Ding$^{1,2}$\cite{HD}, H. Fretwell$^{2}$,
M. Randeria$^3$, and J. C. Campuzano$^{1,2}$}
\address{
         (1) Materials Sciences Division, Argonne National Laboratory,
             Argonne, IL 60439 \\
         (2) Department of Physics, University of Illinois at Chicago,
             Chicago, IL 60607\\
         (3) Tata Institute of Fundamental Research, Mumbai 400005, India\\
         }

\address{%
\begin{minipage}[t]{6.0in}
\begin{abstract}
The self-energy $\Sigma({\bf k},\omega)$, the fundamental function 
which describes the effects of many-body interactions on an electron
in a solid, is usually difficult to obtain directly from
experimental data.  In this paper, we show that by making certain reasonable
assumptions, the self-energy can be directly determined from angle
resolved photoemission data.  We demonstrate this method on data for the
high temperature superconductor $Bi_2Sr_2CaCu_2O_{8+x}$ (Bi2212) in the 
normal, superconducting, and pseudogap phases.
\typeout{polish abstract}
\end{abstract}
\end{minipage}}

\maketitle
\narrowtext

\section{Introduction}

The propagation of an electron in a many-body system is described by the 
Greens function, $G({\bf k},\omega) = 
1 /\left[\omega-\epsilon_{\bf k} -\Sigma({\bf k},\omega)\right]$, 
where $\epsilon_{\bf k}$ is the bare energy of the
electron and the self-energy $\Sigma({\bf k},\omega)$ encapsulates the
effects of many-body interactions.  
A detailed knowledge of $\Sigma({\bf k},\omega)$
is of critical importance in elucidating the microscopic physics of the
system.  If its $\bf k$-dependence is not important,
one can obtain information about $\Sigma$ from a probe
like tunneling, which measures the density of
states given by a $\bf k$-sum of the imaginary part of $G$.  
This was exploited to get a very detailed microscopic understanding 
of strong-coupling electron-phonon superconductors\cite{SCHRIEFFER} 
like lead. In general, though, if $\Sigma$ depends on 
$\bf k$, then momentum averaged 
probes cannot be used to extract the self-energy.

The only truly $\bf k$-resolved probe is angle resolved photoemission
(ARPES).  Under the assumption that the ``sudden" approximation
applies (that is, one can ignore the interaction of the photohole with the
outgoing photoelectron), for quasi-2D systems (since the component of the
momentum perpendicular to the surface is not conserved in the photoemission
process), and assuming only a single initial state (one ``band"), then the
photocurrent can be written in the following form\cite{DING95,NK}
\begin{equation}
I({\bf k},\omega) = C_{\bf k} \sum_{\delta \bf k} \int d\omega'
A({\bf k'},\omega') f(\omega') R(\omega-\omega') + B
\end{equation}
where $C_{\bf k}$ is an intensity prefactor (proportional to the square of
the dipole matrix element between initial and final states), 
$A = (-1/\pi)ImG$ is the single particle spectral
function, $f$ is the Fermi function, and $R$ a Gaussian
energy resolution function (photon monochromator and detector).
The sum $\sum_{\delta \bf k}$ is over a small
window in ${\bf k}$-space due to the finite angular aperture of the detector.
$B$ is the background, which contains extrinsic effects such
as inelastic scattering of the photoelectrons (secondaries).

In this paper, we exploit Eq.~(1) to determine the electron self-energy,
and illustrate this for ARPES data on the high temperature superconductor
Bi2212.  In Section II, we introduce the methodology which is necessary
to extract the self-energy from the data.  In Section III, we discuss the
issue of background subtraction.  In Section IV, various results
are presented for Bi2212 in the normal, superconducting, and pseudogap phases.
Finally, some concluding remarks are offered in Section V.

\section{Methodology}

Let us assume we know $A$.  Given that,
we can easily obtain $\Sigma$.  A Kramers-Kronig transform of $A$ will
give us the real part of $G$
\begin{equation}
ReG(\omega) = P\int_{-\infty}^{+\infty} d\omega'
\frac{A(\omega')}{\omega'-\omega}
\end{equation}
where $P$ denotes the principal part of the integral.
Knowing now both $ImG$ and $ReG$, then
$\Sigma$ can be directly read off from the definition of $G$.
\begin{eqnarray}
Im\Sigma = \frac{ImG}{(ReG)^2 + (ImG)^2} \nonumber \\
Re\Sigma = \omega-\epsilon-\frac{ReG}{(ReG)^2 + (ImG)^2}
\end{eqnarray}

To obtain $ReG$ using Eq.~(2) we need to know $A$
for {\it all} energies.  From ARPES, though, we only know the product of $A$
and $f$.   (While unoccupied states can be studied by inverse photoemission, 
its resolution at present is too poor to be useful for our purposes).
This is not a limitation if an occupied ${\bf k}$-state is being analyzed and
one can either ignore the unoccupied weight or use a simple 
extrapolation for it (except that only $Re\Sigma+\epsilon$ is
determined).  On the other hand, one is usually interested in $k$ vectors near
the Fermi surface.  Therefore a key assumption will have to be made.
We can implement our procedure if we make the assumption 
of particle-hole symmetry, 
$A(\epsilon_{\bf k},\omega)=A(-\epsilon_{\bf k},-\omega)$, 
within the small ${\bf k}$-window centered at ${\bf k}_F$.
Then, $A$ is obtained by exploiting the identity
$A(\epsilon_{\bf k},\omega)f(\omega)+
A(-\epsilon_{\bf k},-\omega)f(-\omega)=A(\epsilon_{\bf k},\omega)$,
which holds even in the presence of the energy resolution integration in 
Eq.~(1).
Note, this can only be invoked at ${\bf k}_F$, and was used
in our past work to remove the Fermi function from ARPES
data \cite{NAT98,NEW}, where it was denoted as the symmetrization 
procedure (note that the ``symmetrized'' data will correspond to the raw 
data for $\omega < \sim -2.2kT$).
Although the particle-hole symmetry assumption is reasonable for small
$|\omega|$ where it can be tested in the normal 
state by seeing whether the ``symmetrized'' spectrum has a maximum at the
Fermi energy ($E_F$), it will almost certainly
fail for sufficiently large $\omega > 0$.
Nevertheless, since we only expect to derive $\Sigma$ for $\omega <0$, 
then the unoccupied spectral weight will affect the result only in two
ways.  The first is through the sum rule $\int d\omega A(\omega) = 1$
which must be used to eliminate the intensity prefactor $C_{\bf k}$ in
Eq.~(1).  From Eq.~(3), we see that violation of the sum rule will
simply rescale $Im\Sigma$, but not $Re\Sigma$ due to the $\omega-\epsilon$
factor.  Our normalization, though, is equivalent to assuming $n_{k_F}$=0.5,
and thus does not involve ``symmetrized'' data.
The second influence comes from the Kramers-Kronig
transformation in Eq.~(2), which is a bigger problem.
Fortunately, the contribution from large
$\omega' > 0$, for which our assumption is least valid, is suppressed
by $1/(\omega'-\omega)$.
Further, for ${\bf k}_F$, $\epsilon_{\bf k}$=0 and
thus $Re\Sigma$ is not plagued by an unknown constant.

Some comments should be made about using real data.
Data noise is amplified in the Kramers-Kronig transformation in Eq.~(2),  
and it is desirable to filter the data.
We have found for our purposes that a wavelet transform works excellently
in this regard, in that it provides smoothed data without any distortion of 
intrinsic spectral features such as the quasiparticle peak.
We employ a `de-noising' algorithm \cite{andras} which transforms the data
into the `wavelet domain' using class 6 complex Daubechies wavelets
\cite{daubechies}.
Then, all wavelet components with absolute values below a certain threshold
are set to zero and the data are transformed back into the signal domain.
The threshold is set at a level which removes all (or most) of the 
noise from the data.
The advantage of using a wavelet transform, over e.g. a  Fourier filter, 
comes from the localised nature of the wavelets in the signal and 
wavelet domain, i.e. the removal of noise from one portion of the data
has no effect on intrinsic features elsewhere.

Moreover, it is desirable to obtain a self-energy which 
is not artificially broadened in $\omega$ due to energy resolution.  
This is handled by deconvoluting the energy resolution out of the data
using a maximum entropy method \cite{dugdale} based on the `Cambridge
Algorithm' \cite{Gull} which we have found to be quite stable.
Here, the entropy of a distribution is defined to be 
$S=-\sum_n p_{n} \ln p_{n}$, where $p_{n}$ is the intensity at point $n$. 
The algorithm locates the solution with maximum entropy, subject to its
being consistent with the data when convoluted with the experimental 
resolution. 
Consistency testing is done using the $\chi^{2}$ statistic (C),
C $= \sum_n (D_{n}-F_{n}*R)^2/\sigma_{n}^2$, where $D_{n}$ are the data,
$F_{n}*R$ is the solution ($F_{n}$) convoluted with the resolution 
function (R), and $\sigma_{n}^2$ is the variance of datum $n$.
Since a completely flat solution has maximum entropy, the 
algorithm selects the smoothest (`deconvoluted') solution consistent with 
the original data and
should only generate structures that are demanded by the data, i.e. those
which are above the noise.
To minimize the effects of the resolution, we use a high resolution data 
set ($\sigma$=7.5meV, FWHM=18meV) in the low binding energy range, and combine
this with a
lower resolution data set ($\sigma$=15meV, FWHM=35meV) to extend the spectrum
out to higher
binding energy (this takes advantage of the fact that sharp
spectral structures only appear at low binding energies).
The effects of broadening due to the finite momentum window can be minimized
by looking at regions of the Brillouin zone where the
dispersion is weak, which is the case considered here (our momentum
window has a radius $0.045\pi/a$).  This will be less of an issue
when considering data from new high resolution detectors currently becoming
available, where the momentum window can be smaller in area by a factor 
of 25 or more.

\section{Background Subtraction}

We now illustrate our method by using data from the high temperature
superconductor Bi2212.  We choose this material because of its obvious 
interest to the condensed matter physics community, the novel 
electronic phases it 
exhibits as a function of doping and temperature, its lack of 
dispersion along the c-axis which justifies the 2D approximation implicit in 
Eq.~(1) \cite{DING96}, and our own strong familiarity with its spectra.
On the downside, there is the background issue implicit in Eq.~(1).
Looking at the ARPES spectra over a large binding energy range,
we see that the near-$E_F$ spectral features of interest to us
ride on top of a large background. Not only is it too large to be
ascribed entirely to the incoherent part of the spectral 
function $A$, the ratio of the spectral peak to the background
changes with photon energy implying that most if not all of the background is
extrinsic.
This is supported by the fact that the magnitude of the background is
sensitive to the photon incident angle and polarization.
Moreover, for $\bf k$ vectors where the spectral 
peak has dispersed through $E_F$, this background is still present.
It is flat in energy and extends all the way to $E_F$ above $T_c$,
but is gapped in the superconducting state. Taking all of the above facts into 
account, the likely source of the background is scattering from 
other $k$ vectors outside the nominal momentum window, probably due to surface
roughness and/or the incommensurate nature of the Bi2212 superstructure.

There are a number of potential ways in which to subtract this 
background.  An ideal way if one is along a symmetry axis (seldom the case)
is to subtract 
data from perpendicular photon polarizations so as to recover that part of the 
signal which obeys the appropriate dipole selection rules.  In practice, 
one is usually limited to subtracting data 
from two perpendicular $k$ vectors since the polarization is fixed.  
Moreover, the finite diameter of the momentum window, possible sample 
alignment errors, and the enhancement of noise due to subtracting two 
data sets, limit the effectiveness of this method.  Another possibility is to 
subtract data from an unoccupied $k$ vector under the assumption that it 
is all background.  The obvious problem here, besides the above mentioned 
amplification of data noise due to subtracting two data sets, is the 
strong variation of the dipole matrix elements in the Brillouin 
zone\cite{BANSIL} that can also act to modulate the intensity of the 
background from one $k$ vector to the next.

Because of this, we have instead explored models which capture 
the essense of the observed background, in particular
a step-edge (flat) background
and a ``Shirley'' background\cite{SHIRLEY}.  The
latter is of the form\cite{BACK}
\begin{equation}
I(\omega) = P(\omega) + c_{Sh}\int_{\omega}^{\infty} d\omega' P(\omega')
\end{equation}
where $I$ is the total intensity and $P$ that due to primary electrons (thus,
one solves for $P$ by simple matrix inversion).
Although the step-edge background looks like the ARPES intensity seen for 
unoccupied $\bf k$ states (hence its motivation), it has the disadvantage of
having three 
adjustable parameters (its height, and the position and width of its leading
edge).
Despite the fact that the Shirley background is designed to model secondary
emission, an unlikely source of the background\cite{BACK,EELS}, it is similar
to the step-edge background, has the advantage of only one adjustable
parameter, and has been used extensively in previous
treatments\cite{OLSON,BACK}.

To implement the background subtraction, the high energy tail of the data 
is fit to a constant plus a Lorentzian, and then $c_{Sh}$ in Eq.~(4) is 
varied such that this constant becomes zero.  This results in a smaller 
background than simply forcing the intensity to be all background beyond some
energy.  This is done for data up to 0.5 eV where a minimum is seen in the
spectrum, since beyond this, the spectrum rises and thus the ``tail'' becomes
completely buried under emission associated with the main valence band.
For energies beyond 0.5 eV, we assume this Lorentzian tail when
performing the integral (to infinity) in Eq.~(2) (real, not fitted, data is
used below 0.5 eV, of course).  The purpose of this procedure is to avoid
artificially forcing
$Im\Sigma$ to zero at some cut-off (a power law tail is not used because 
the resulting integral would not be convergent).  Once this background is
subtracted, then the data are symmetrized (by adding the data at positive 
and negative energies), normalized (by invoking the sum rule), then 
Kramers-Kronig transformed, which is done analytically by assuming $A$ 
to be linear between data points (the Lorentzian tail beyond 0.5 eV has an
analytic transform, of course).
A similar procedure is used for the step-edge
background.  The height of the step is determined by fitting the
high energy data to a constant plus a Lorentzian.  In the superconducting 
state, the position and width of the step's
leading edge is determined by fitting the low energy data to a Fermi 
function (whose ``chemical potential'' is the position and whose 
``temperature'' is the width of the step) plus a Gaussian (modeling
the spectral peak).  In the normal state, the step-edge background simply 
reverts to a constant in the symmetrized data and so no low energy 
modeling is necessary.

\section{Results}

\begin{figure}
\epsfxsize=3.4in
\epsfbox{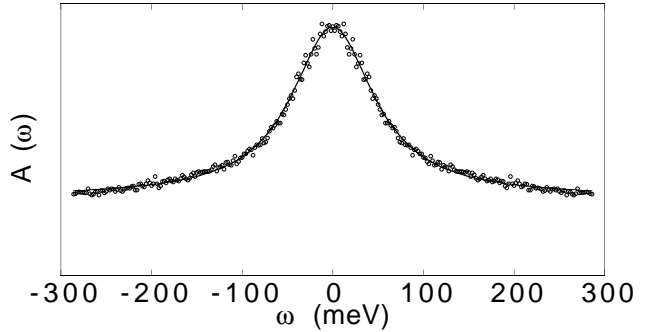}
\vspace{0.5cm}
\caption{
Symmetrized spectrum for overdoped Bi2212 ($T_c$=72K) at $T$=80K at the
$(\pi,0)-(\pi,\pi)$ Fermi crossing,
with the line a fit to a constant plus Lorentzian.  For visual purposes,
it is shown for $\omega > 0$, though we expect reliable information
only for $\omega < 0$.  This applies to all the figures.}
\label{fig1}
\end{figure}

In Fig.~1, we show symmetrized data at the $(\pi,0)-(\pi,\pi)$ 
Fermi crossing for a $T_c$=72K overdoped sample at $T$=80K, and thus in 
the normal state.  Note that the spectral peak is centered at zero energy,
consistent with being at $k_F$ with the zero of energy at $E_F$.
The line is a fit to a Lorentzian 
plus a constant (flat background), and is an excellent representation 
of the data (with a HWHM of 55 meV).
This Lorentzian spectral shape at ${\bf k}$ is sufficiently
broad to make the quasiparticle ill-defined, but may seem unusual given 
the supposedly expected marginal Fermi liquid form\cite{MFL}.
We have always found Lorentzian fits to work well in the vicinity of the
$(\pi,0)$ point in the normal state\cite{OLSON2}.  
Moreover, in Bi2201, where the normal state can be accessed over a 
large temperature range, we again find equally good Lorentzian fits even 
at low temperatures.  The difference from optical conductivity 
data\cite{TIMUSK}, which do indicate a marginal Fermi liquid form, 
may be resolved by noting that the region near $(\pi,0)$
makes little contribution to the in-plane transport due to
the flat dispersion.  In fact, near the direction $(0,0)-(\pi,\pi)$, a case 
has been made that a marginal Fermi liquid lineshape can adequately 
describe the data if a background subtraction similar to what we employ
here is done \cite{OLSON,BACK}.  This points to the possibility of a
variation of the momentum dependence of $\Sigma$ along the Fermi surface,
which our method can in 
principle explore with the advent of new detectors with improved momentum 
resolution.  Finally, we note that if we restrict away from
small energies, a constant plus a power law fits the data as well as a 
constant plus a Lorentzian.  Typically, the (negative) power, $\alpha$, is
such that $|\alpha|<1$ (smaller for smaller doping), which would be consistent 
with a non Fermi liquid line shape \cite{ANDERSON,MISRA}.  The advantage of the 
Lorentzian is that it goes through all the data, not just the higher 
energy part, though this may be fortuitous if part of the ``background'' 
turns out to be intrinsic (the power law fit has the potential advantage 
of a smaller constant background than the Lorentzian fit).
A power law tail would also be divergent in Eq.~(2), and thus would have
to be cut-off (how to do this is not clear, since the tail is buried under
the main valence band emission).  This issue will hopefully be 
resolved in the future by doing a detailed analysis of the spectra as a 
function of photon energy, photon incident angle, and polarization to 
determine how much of the ``background'' is truly extrinsic.  Once this 
is achieved, a closer representation of the true self-energy can be obtained.

\begin{figure}
\epsfxsize=3.4in
\epsfbox{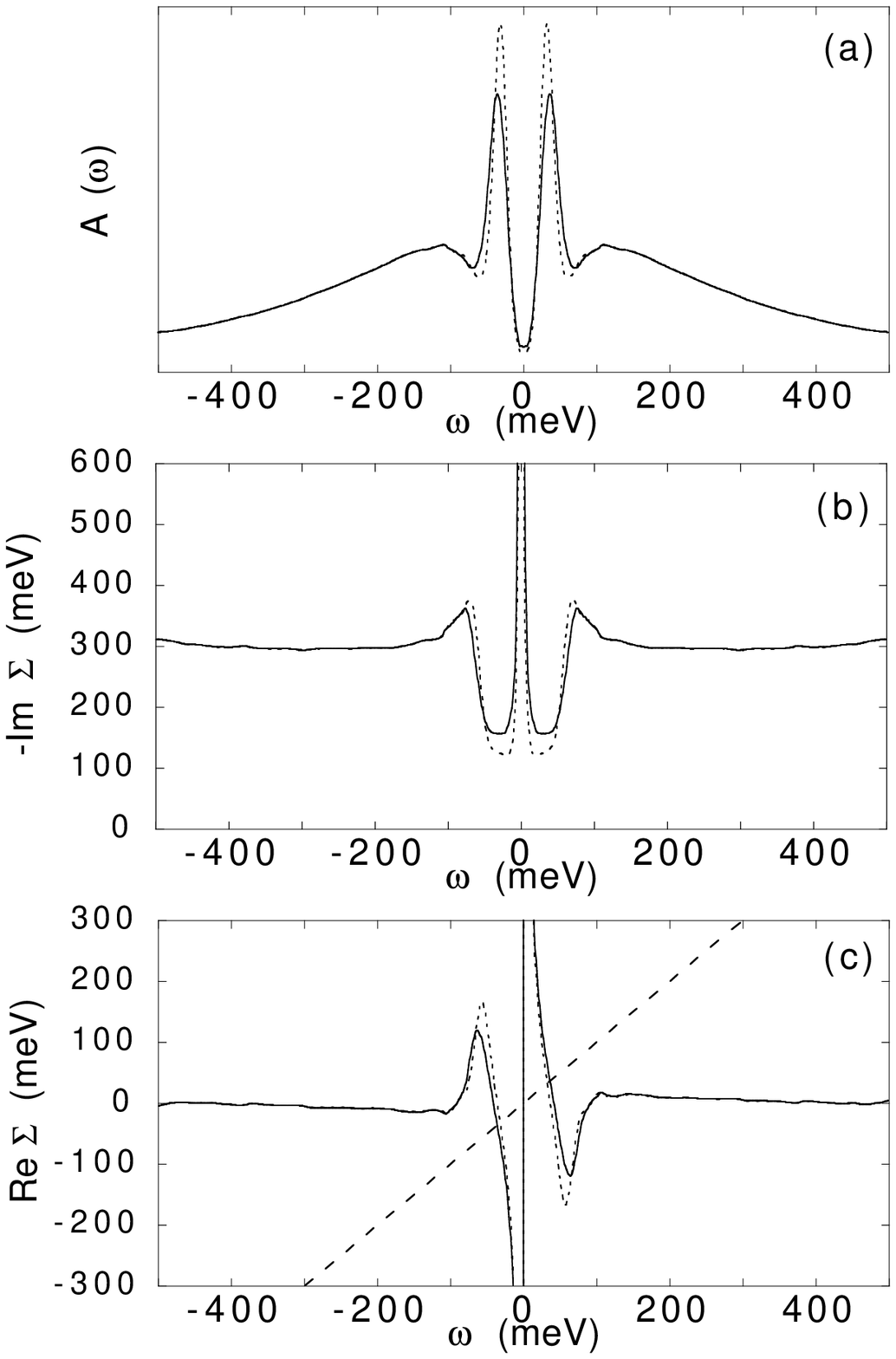}
\vspace{0.5cm}
\caption{
\label{fig2}
(a) Symmetrized spectrum (smoothed and Shirley subtracted) for overdoped Bi2212
($T_c$=87K) at $T$=14K at $(\pi,0)$ with (dotted line) and without
(solid line) resolution deconvolution.
The resulting $Im\Sigma$ and $Re\Sigma$ are shown in (b) and (c).
The dashed line in (c) determines the condition $Re\Sigma=\omega$.}
\end{figure}

In Fig.~2(a), we show $T$=14K symmetrized data for a
$T_c$=87K overdoped sample at the $(\pi,0)$ point 
(data of Ref.\onlinecite{COLL}).
We note the important differences in this superconducting state
spectrum, compared with the normal state spectrum
in Fig.~1, due to the opening of the superconducting gap, with
the appearance of a sharp quasiparticle peak displaced from $E_F$ 
by the superconducting gap, followed by a spectral dip, then by a 
``hump''\cite{DESSAU} at higher binding energies 
(where the normal and superconducting state spectra 
coincide \cite{NK,COLL}).  This unusual dip-hump structure is only
seen near $(\pi,0)$.
The resulting $\Sigma$ is shown in Fig.~2(b) and (c).
At high binding energies, one obtains a constant $Im\Sigma$ 
as expected from the Lorentzian behavior above $T_c$ in Fig.~1.  Note the 
very large value ($\sim$ 300 meV), much larger than that implied by Fig.~1
(this is verified by the normal state spectrum, which has a much larger 
HWHM than the normal state spectrum of Fig.~1).  That is, the magnitude 
of $Im\Sigma$ strongly increases with reduced doping.
Near the spectral dip, $Im\Sigma$ has a small peak followed by a sharp 
drop, which we had earlier inferred\cite{COLL} from
the spectral shape guided by fits to the data\cite{ND}.
This behavior is expected if the electrons 
are interacting with a spectral distribution gapped by 
2$\Delta$ in the superconducting state together with a sharp 
collective mode inside the 2$\Delta$ gap. 
The current results fully confirm the collective
mode explanation proposed in Refs.~\onlinecite{COLL} and \onlinecite{ND}.

Despite this sharp drop below 70 meV,
$Im\Sigma$ remains quite large at low frequencies.
That is, the quasiparticle peak is not resolution limited.  
It's flat behavior ($\omega^6 \sim \omega^7$) between
20 and 60 meV is consistent with the 
$T^6$ dependence of the quasiparticle peak width noted
in Ref.~\onlinecite{NEW}.
Then, below 20 meV, there is a narrow spike in $Im\Sigma$.  
This is the imaginary part of the 
BCS self-energy, $\Delta^2/(\omega+i0^+)$, which
kills the normal state pole at $\omega$=0.  The resulting 1/$\omega$
divergence of the real part $Re\Sigma$, which creates new poles at 
$\pm\Delta$=32meV, is easily seen in Fig.~2(c).
This is followed by a strong peak in $Re\Sigma$ near the spectral dip 
energy, which follows from the Kramers-Kronig transformation of the 
sharp drop in $Im\Sigma$. The strong peak in $Re\Sigma$ explains why the low
energy peak in $A$ is so narrow despite the large value of $Im\Sigma$.
The halfwidth of the spectral peak is given by $\Gamma=Im\Sigma/Z$ where
$Z=1-\partial Re\Sigma/\partial \omega$ (the inverse of the quasiparticle
residue).  In the vicinity of the
spectral peak, $Z$ is large ($\sim$9), giving a $\Gamma$ of $\sim$14 meV.
We note, though, that $\Gamma$ is still quite sizeable, and thus the peak 
is not resolution limited.  As the peak is dispersionless near 
$(\pi,0)$\cite{COLL}, this width is unlikely to be due to momentum resolution,
which was verified by simulation.
One could ask if it were due to an improper energy resolution deconvolution.
This is highly unlikely, which was also checked by simulation.  For instance,
if one fits the zero energy spike in 
$Im\Sigma$ to a constant plus a Lorentzian, the resulting Lorentzian is 
extremely narrow (with a HWHM of 2 meV).

It is crucial to understand the extent to which our results
for $\Sigma$ depend upon the choice of various background functions.
In Fig.~3, we compare $Im\Sigma$ (as in Fig.~2(b)) for three 
different background choices: Shirley, step-edge, and
no subtraction at all (for the last case, the spectrum is simply chopped 
off at 0.5 eV binding energy, and thus no Lorentzian tail).
It is reassuring that all three results are qualitatively similar
(at higher binding energies, the unsubtracted case decays to zero 
because of the cut-off). There are some interesting quantitative 
differences of the step-edge background from the other two, in particular 
the step-like drop in $Im\Sigma$ is more pronounced (resulting in a 
much more pronounced peak in $Re\Sigma$). This behavior is not very 
sensitive to the choice of the leading edge position and width of the step-edge
background, and the result is quantitatively close to the theory of 
Ref.~\onlinecite{ND}.  In all cases, $Im\Sigma$ is quite large at low 
energies, consistent with a quasiparticle peak which is not resolution 
limited.

\begin{figure}
\epsfxsize=3.4in
\epsfbox{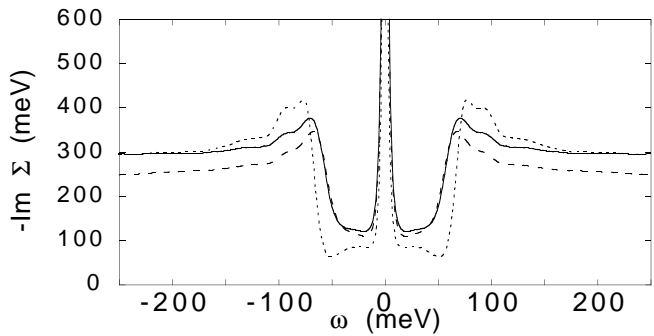}
\vspace{0.5cm}
\caption{
\label{fig3}
$Im\Sigma$ as in Fig.~2(b) (with resolution deconvolution), but for three
different background subtractions:  Shirley (solid line), step-edge
(dotted line),
and no subtraction with a cut-off at 0.5eV (dashed line).}
\end{figure}

We have also looked at data from a $T_c$=85K underdoped Bi2212 sample 
(data of Ref.~\onlinecite{NAT98}).  
Below $T_c$ we find behavior quite similar to that of Fig.~2.
Of more interest in this case is the so-called pseudogap phase, 
where a gap is seen in the spectral function above
$T_c$\cite{DING,LOESER}.
In Fig.~4(a), we show $T$=95K symmetrized data at the 
$(\pi,0)-(\pi,\pi)$ Fermi crossing.  One again sees (Fig.~4(b)) a
peak in $Im\Sigma$ at $\omega$=0, but it is broadened relative to
that of the superconducting state, and the corresponding divergence of
$Re\Sigma$ (Fig.~4(c)) is smeared out.  Such behavior would be consistent with 
replacing the BCS self-energy $\Delta^2/(\omega+i0^+)$ by 
$\Delta^2/(\omega+i\Gamma_0)$. We have recently shown that such a 
self-energy gives a good description of low energy data\cite{NEW}, 
and can be motivated by considering the presence
of pair fluctuations above $T_c$.  In fact, the $\Sigma$ of
Fig.~4 looks remarkably similar to the simple form proposed in
Ref.~\onlinecite{NEW}, even over a large binding energy range.
Note from Fig.~4 that although the equation
$\omega-Re\Sigma(\omega)=0$ is still satisfied at $|\omega| \sim \Delta$, 
$Im\Sigma/Z$ is so large that the spectral peak is strongly broadened
in contrast to the sharp peak seen below $T_c$.  Actually, to a good
approximation, the spectral function is essentially the inverse of $Im\Sigma$
in the range $|\omega| < \sim 2\Delta$.
We can also contrast this case with data taken
above $T^*$, the temperature at which the pseudogap ``disappears".  In
that case, the spectrum is featureless, and the peak in $Im\Sigma$
is strongly broadened.  As the doping increases, this peak in $Im\Sigma$
disappears.
Further doping causes a depression in $Im\Sigma$ to develop around
$\omega=0$, indicating a crossover to more Fermi liquid like behavior.

\begin{figure}
\epsfxsize=3.4in
\epsfbox{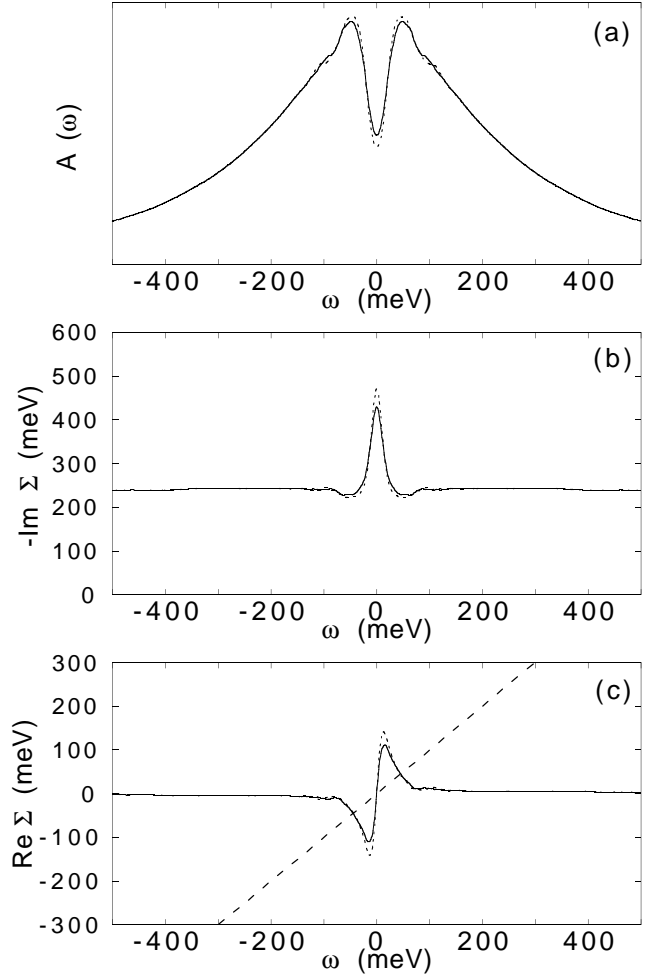}
\vspace{0.5cm}
\caption{
\label{fig4}
(a) Symmetrized spectrum (smoothed and Shirley subtracted) for underdoped Bi2212
($T_c$=85K)
at $T$=95K (pseudogap phase) at the $(\pi,0)-(\pi,\pi)$ Fermi crossing with
(dotted line) and without (solid line) resolution deconvolution.
The resulting $Im\Sigma$ and $Re\Sigma$ are shown in (b) and (c).  The dashed
line in (c) determines the condition $Re\Sigma=\omega$.}
\end{figure}

\section{Conclusions}

In conclusion, we have proposed a method for determining the
self-energy $\Sigma({\bf k},\omega)$ from ARPES data.  
Although several important assumptions have to be made (particle-hole symmetry,
background subtraction), the method has the advantage 
that one can directly determine $\Sigma$,
rather than attempt to guess it by fitting the data\cite{NEW,ND,TT}.   
Given the wealth of information one can obtain,
we expect this procedure to be very useful in elucidating the 
microscopic physics of solids, particularly low dimensional strongly 
correlated systems where many controversies exist. Specifically, we find
a non-trivial 
frequency dependence of $\Sigma$ in the superconducting and pseudogap
phases of the high temperature cuprate superconductors, which puts 
strong constraints on the microscopic theory for these materials.

\acknowledgments

We thank Yuri Vilk for a key suggestion.  This work was supported by the U. S.
Dept. of Energy,
Basic Energy Sciences, under contract W-31-109-ENG-38, the National 
Science Foundation DMR 9624048, and
DMR 91-20000 through the Science and Technology Center for
Superconductivity.

\end{document}